\documentclass[prb, aps, twocolumn]{revtex4}
\usepackage{graphicx}
\usepackage{bm}
\usepackage{amssymb}
\usepackage{amsmath}
\usepackage{amsfonts}
\usepackage{amsopn} 
\DeclareMathOperator{\trace}{tr}

\newcommand {\bra} [1] {\langle #1 |}
\newcommand {\ket} [1] {| #1 \rangle}
\newcommand {\pd} [2] {\frac{\partial #1}{\partial #2}}
\newcommand {\td} [2] {\frac{d #1}{d #2}}
\begin{document}
\title{On the nature of steady states of spin distributions in the presence of spin-orbit interactions}
\author{Dimitrie Culcer}
\author{R. Winkler}
\affiliation{Advanced Photon Source,
Argonne National Laboratory, Argonne, IL 60439.}
\affiliation{Northern Illinois University, De Kalb, IL 60115.}
\begin{abstract}
In the presence of spin-orbit interactions, the steady state
established for spin distributions in an electric field is
qualitatively different from the steady state for charge
distributions. This is primarily because the steady state
established for spin distributions involves spin precession due to
spin-orbit coupling. We demonstrate in this work that the spin
density matrix in an external electric field acquires two
corrections with different dependencies on the characteristic
momentum scattering time. One part is associated with conserved
spins, diverges in the clean limit and is responsible for the
establishment of a steady-state spin density in electric fields.
Another part is associated with precessing spins, is finite in the
clean limit and is responsible for the establishment of spin
currents in electric fields. Scattering between these
distributions has important consequences for spin dynamics and
spin-related effects in general, and explains some recent puzzling
observations, which are captured by our unified theory.
\end{abstract}
\date{\today} \maketitle

\section{Introduction}

Spin-orbit interactions are frequently the most important factor
determining spin dynamics in solids. From a technological
perspective, novel physical phenomena that may lead to improved
memory devices and advances in quantum information processing have
been shown to be intimately related to spin-orbit interactions.
\cite{zut04} Practical applications usually rely on generating and
maintaining a spin polarization, and in this context spin-orbit
interactions can play a constructive or a destructive role. In an
external electric field spin precession gives rise to steady-state
spin densities\cite{vor79, ivc78, lev85, ede90, aro91, cha01,
ino03, eng07, kato04b, gan04, sil04} and spin currents,
\cite{eng06, dya71, hir99, mur03, kato04, wun05, sih05, val06,
liu06, ste06, gan01, sin04, ada05, shi06, ras04, zha05, wang06,
she05, han06, kha06, shy06, dam04, schl05, jin06, li04, ino04,
sch02, dim05, mal05, eng05, nik06, ble06, kro06, zar06, mis04,
nik04, win06, sug06} yet spin precession is often the leading
cause of spin polarization decay. \cite{dya72, dya86, pik84,
ave02} These facts suggest that spin precession plays a nontrivial
role in the establishment of a steady state for spin distributions
in electric fields.

Spin-orbit interactions are present in the band structure and in
potentials due to impurity distributions. Spin-orbit coupling is
in principle always present in impurity potentials and gives rise
to skew scattering, which leads to the extrinsic anomalous Hall
effect and the extrinsic spin-Hall effect. \cite{ono06} Band
structure spin-orbit coupling may arise from the inversion
asymmetry of the underlying crystal lattice \cite{dre55} (bulk
inversion asymmetry), from the inversion asymmetry of the
confining potential in two dimensions \cite{ras84} (structure
inversion asymmetry), and may be present also in inversion
symmetric systems. \cite{lut56} Band structure spin-orbit coupling
gives rise to spin precession, is the cause of magnetic anisotropy
in magnetic semiconductors, \cite{mac06} and causes spin flips
even in the course of elastic scattering by spin-independent
potentials. \cite{pik84, ell54} Band structure spin-orbit
interactions in spin-1/2 electron systems can always be
represented by a Zeeman-like Hamiltonian $H = (1/2) \,
\bm{\sigma}\cdot\bm{\Omega}_{\bm{k}}$ describing the interaction
of the spin with an effective wave vector-dependent magnetic field
$\bm{\Omega}_{\bm{k}}$. An electron spin at wave vector ${\bm k}$
precesses about this field with frequency $\Omega_{\bm{k}}/\hbar
\equiv |{\bm \Omega_{\bm{k}}}|/\hbar$ and is scattered to a
different wave vector within a characteristic momentum scattering
time $\tau_p$. Throughout this paper we assume that
$\varepsilon_F\tau_p/\hbar \gg 1$, where $\varepsilon_F$ is the
Fermi energy, which is tantamount to assuming that the carriers'
mean free path is much larger that their de Broglie wavelength.
Within this range, the relative magnitude of the spin precession
frequency $\Omega_{\bm{k}}$ and inverse scattering time $1/\tau_p$
define three qualitatively different regimes. In the ballistic
(clean) regime no scattering occurs and the temperature tends to
absolute zero, so that $\varepsilon_F\tau_p \rightarrow \infty$
and $\Omega_{\bm{k}}\tau_p/\hbar \rightarrow \infty$. The weak
scattering regime is characterized by fast spin precession and
little momentum scattering due to, e.g., a slight increase in
temperature, yielding $\varepsilon_F\tau_p/\hbar \gg
\Omega_{\bm{k}} \, \tau_p/\hbar \gg 1$. In the strong momentum
scattering regime $\varepsilon_F\tau_p/\hbar \gg 1 \gg
\Omega_{\bm{k}} \, \tau_p/\hbar $.

In this paper, we will be concerned with the interplay of external
fields, scattering processes, and band structure spin-orbit
interactions in establishing a steady state for spin
distributions. We will concentrate on electrons in uniform
electric fields. In charge transport, the steady state is
characterized by a nonequilibrium correction to the density matrix
that is divergent in the clean limit, indicating a competition
between the electric field, accelerating charge carriers, and
scattering, which inhibits their forward motion. On the other
hand, nonequilibrium corrections that arise as a result of band
structure spin-orbit coupling in crystal Hamiltonians
re\-pre\-sent a different kind of interplay between the electric
field and scattering processes. Firstly, the spin-orbit splitting
of the bands gives rise to spin-dependent scattering even from
spin-independent potentials. Secondly, the presence of spin
precession causes the steady state established for spin
distributions to be highly nontrivial. We will demonstrate that
the density matrix contains a contribution due to precessing spins
and one due to conserved spins, and the steady states established
for each of these are qualitatively different. Steady state
corrections $\propto \tau_p$, which diverge in the clean limit,
are associated with the \emph{absence} of spin precession and give
rise to spin densities in external fields. \cite{vor79, ivc78,
lev85, ede90, aro91, cha01, ino03, eng07, kato04b, gan04, sil04}
Steady state corrections independent of $\tau_p$, which are finite
in the clean limit, are associated with spin precession and give
rise to spin currents in external fields. \cite{eng06, dya71,
hir99, mur03, kato04, wun05, sih05, val06, liu06, ste06, gan01,
sin04, ada05, shi06, ras04, zha05, wang06, she05, han06, kha06,
shy06, dam04, schl05, jin06, li04, ino04, sch02, dim05, mal05,
eng05, nik06, ble06, kro06, zar06, mis04, nik04, win06, sug06}
Scattering between these two distributions induces significant
corrections to steady-state spin densities and spin currents.

In research on electrical generation of spin densities and
currents two rather different approaches are employed. Linear
response theories based on Green's functions \cite{ede90, ino03,
han06, dam04, ino04, dim05, mal05, ble06, nik06, sch02, ono06}
provide a diagrammatic interpretation of spin-related phenomena in
electric fields. Such theories usually provide reliable results,
but the physical picture is not always evident. Theories based on
a kinetic equation for the density matrix \cite{lev85, eng07,
aro91, mis04, kha06, shy06, eng05} tend to be transparent.
However, it is often difficult, in such theories, to determine
from the outset all the terms that play an important role in the
processes under study. Correspondences between the two approaches
mentioned were identified in a recent paper by Sinitsyn
\textit{et~al}. \cite{nik07} The formalism used in the present
paper, although relying on a density matrix, is equivalent to Kubo
linear response theory.

We consider large, uniform systems, working in momentum space
without making semiclassical approximations (by which we
understand approximations pertaining to simultaneous consideration
of a particle's position and momentum.) The role of scattering in
spin-related effects is an issue that is currently under intense
investigation. In our general formalism, the terms responsible for
scattering are derived rigorously beginning with the quantum
Liouville equation for the density operator. One particular issue
at the center of current debate is the physical interpretation of
vertex corrections in spin-related phenomena. In charge transport,
it is well known that vertex corrections decrease the weight of
small-angle scattering. We demonstrate that, in spin transport,
scattering phenomena have two main consequences. Firstly, the
driving term in the equation for the spin density acquires a
contribution due to spin-dependent scattering. Secondly,
scattering between the distributions of conserved and precessing
spins enters the equations for these two distributions. Our work
suggests that both of these processes are contained in vertex
corrections.

In a recent related article, \cite{dim07} we focused on the spin
current response of a semiconductor to an electric field, showing
that spin currents are not restricted to the spin-Hall effect and
dwelling briefly on the distinction between nonequilibrium spin
currents and spin densities excited by electric fields. In this
article, we concentrate on the steady state and on the difference
between the steady state established for spin distributions and
that established for charge distributions. We investigate aspects
of the steady state for spin distributions that arise as a result
of spin precession and have no analog in charge distributions. We
provide a detailed analysis of the distributions of conserved and
precessing spins, and attempt to shed light on the physical
content of vertex corrections in systems with spin-orbit
interactions. In the process, we give a detailed exposition of the
underlying theory, which was also employed in our recent work.
\cite{dim07}

The outline of this paper is as follows. In the next section we
start from the quantum Liouville equation and derive an equation
describing the time evolution of the electron density matrix
including the full scattering term in the first Born
approximation. Section III is devoted to the complex case of
steady states in the presence of band structure spin-orbit
coupling, demonstrating the existence of two spin distributions
and their subtle interplay. Analytical expressions are given for
general scattering as far as possible, as well as exact solutions
for short-range impurity scattering. We close with a summary of
our findings.

\section{Time evolution of the density operator}
We consider a system of non-interacting spin-1/2 electrons. The
electrons are represented by a one-particle density operator $\hat
\rho$. The expectation value of an observable represented by a
Hermitian operator $\hat O$ is given by $\trace (\hat\rho\hat
O)$, which motivates us to study the density operator $\hat\rho$ in
detail. The dynamics of $\hat \rho$ are described by the quantum
Liouville equation, which is projected onto a set of states of
definite wave vector (in which the matrix elements of $\hat \rho$
form the density matrix), and an equation is obtained for the time
evolution of the density matrix. This equation is valid for any
elastic scattering in the first Born approximation. In this
article we discuss systems with long mean free paths and do not
consider diffusion terms explicitly. Consequently, semiclassical
approximations are not necessary, and the method employed in the
present work is equivalent to the Kubo linear response formalism.

\subsection{Quantum Liouville equation}

The quantum Liouville equation satisfied by $\hat \rho$ is
\begin{equation}
\td{\hat\rho}{t} + \frac{i}{\hbar} \, [\hat H + \hat U, \hat \rho]
= 0.
\end{equation}
The Hamiltonian $\hat H$, considered here in the framework of the
envelope function approximation (EFA), contains contributions due
to the kinetic energy and spin-orbit coupling. The effect of the
lattice-periodic potential of the ions is taken into account
through a replacement of the carrier mass by the effective mass.
We shall henceforth refer to $\hat H$ as the EFA Hamiltonian.
Scattering is introduced into the system through the potential
$\hat U$, which may be due to impurities, phonons, surface
roughness, or other perturbations. In this article we focus on
impurity scattering, as the effects that are discussed are
frequently observed at very low temperatures, where the role of
phonon scattering may be neglected.

The Liouville equation is projected onto a set of time-independent
states of definite wave vector $\{ \ket{{\bm k}s} \}$, which are
not assumed to be eigenstates of the EFA Hamiltonian $\hat H$. The
matrix elements of $\hat \rho$ in this basis will be written as
$\rho_{{\bm k}{\bm k}'} \equiv \rho^{ss'}_{{\bm k}{\bm k}'} =
\bra{{\bm k}s} \hat\rho \ket{{\bm k}'s'}$, with corresponding
notations for the matrix elements of $\hat H$ and $\hat U$. Spin
indices will not be shown explicitly in our subsequent derivation,
the quantities $\rho_{{\bm k}{\bm k}'}$, $H_{{\bm k}{\bm k}'}$,
and $U_{{\bm k}{\bm k}'}$ being treated as matrices in spin space.
$\rho_{{\bm k}{\bm k}'}$ is referred to as the density matrix.
With our choice of basis functions of definite wave vector, matrix
elements of the Hamiltonian $H_{{\bm k}{\bm k}'} = H_{{\bm k}} \,
\delta_{{\bm k}{\bm k}'}$ are diagonal in ${\bm k}$. However, if
the EFA Hamiltonian contains spin-orbit coupling terms, the matrix
elements $H_{{\bm k}}$ are generally off-diagonal in spin space.
Matrix elements of the scattering potential $U_{{\bm k}{\bm k}'}$
are off-diagonal in ${\bm k}$. Matrix elements diagonal in ${\bm
k}$ in the scattering potential would lead to a redefinition of
$H_{{\bm k}}$, which is analogous, in Green's function formalisms,
to the offset introduced by the real part of the self energy.
Scattering is assumed elastic and, given that we will work in the
first Born approximation, the matrix $U_{{\bm k}{\bm k}'}$ is
assumed diagonal in spin. (We thus do not take into account
so-called skew scattering terms, which require terms of third
order in $U_{{\bm k}{\bm k}'}$ as well as explicit inclusion of
spin-orbit coupling in the scattering potential, and which were
studied in a recent work. \cite{eng05}) As mentioned above, we
focus in this article on impurity scattering. The impurities are
assumed uncorrelated and our normalization is such that $\bra{{\bm
k}s}\hat U\ket{{\bm k}'s'}\bra{{\bm k}'s'}\hat U\ket{{\bm k}s} =
n_i |U_{{\bm k}{\bm k}'}|^2 \delta_{ss'}$, where $n_i$ is the
impurity density. $U_{{\bm k}{\bm k}'}$ refers therefore to the
matrix elements of the potential due to a single impurity.
Explicit expressions for these matrix elements for a screened
Coulomb potential in two and three dimensions are given in
Appendix A.

$\rho_{{\bm k}{\bm k}'}$ is divided into a part diagonal in ${\bm
k}$ and a part off-diagonal in ${\bm k}$, given by $\rho_{{\bm
k}{\bm k}'} = f_{{\bm k}} \, \delta_{{\bm k}{\bm k}'} + g_{{\bm
k}{\bm k}'}$, where, in $g_{{\bm k}{\bm k}'}$, it is understood
that ${\bm k} \ne {\bm k}'$. The quantum Liouville equation can be
broken down into equations for $f_{{\bm k}}$ and $g_{{\bm k}{\bm
k}'}$
\begin{subequations}
\begin{eqnarray}
\td{f_{{\bm k}}}{t} + \frac{i}{\hbar} \, [H_{{\bm k}}, f_{{\bm
k}}] & = & - \frac{i}{\hbar} \, [\hat U, \hat g]_{{\bm k}{\bm k}} ,
\\ [1ex] \td{g_{{\bm k}{\bm k}'}}{t} +
\frac{i}{\hbar} \, [\hat H, \hat g]_{{\bm k}{\bm k}'} &
 = & - \frac{i}{\hbar} \, [\hat U, \hat f + \hat g]_{{\bm k}{\bm
k}'} \label{eq:g}.
\end{eqnarray}
\end{subequations}
In the first Born approximation the solution to Eq.\ (\ref{eq:g})
can be written as
\begin{equation}
g_{{\bm k}{\bm k}'} = - \frac{i}{\hbar} \, \int_0^\infty dt'\,
 e^{- i \hat H t'} \left[\hat U, \hat f (t - t') \right] e^{i
\hat H t'}|_{{\bm k}{\bm k}'}.
\end{equation}
In order to shorten the equations, factors of $\hbar$ that appear
in the time evolution operators will be omitted, i.e., $e^{i \hat
H t'} \equiv e^{i \hat H t'/\hbar}$. These factors will be
restored in the final results. Since $\varepsilon_F\tau_p/\hbar
\gg 1$, we shall expand $\hat f(t - t')$ in the time integral
around $t$ and, noting that terms beyond $\hat f(t)$ are of higher
order in the scattering potential, we shall only retain the first
term, $\hat f(t)$. The equation for $f_{{\bm k}}$ then becomes
\begin{subequations}\label{eq:FermiJfk}
\begin{equation}\label{eq:Fermi}
\td{f_{{\bm k}}}{t} + \frac{i}{\hbar} \, [H_{{\bm k}}, f_{{\bm
k}}] + \hat J (f_{{\bm k}}) = 0, \end{equation} in which the
scattering term $\hat J (f_{{\bm k}})$ is given by
\begin{equation}
\label{eq:Jfk} \hat J (f_{{\bm k}}) = \frac{1}{\hbar^2}
\,\int_{0}^\infty dt'\, \left[\hat U, e^{- i \hat H t'} \left[\hat U, \hat f
(t) \right] e^{ i \hat H t'} \right]_{{\bm k}{\bm k}}.
\end{equation}
\end{subequations}
The integral over time in Eq.\ (\ref{eq:Jfk}) can be performed by
inserting a regularizing factor $e^{- \eta t'}$ and letting $\eta
\rightarrow 0$ subsequently. We remark that, for potentials
diagonal in spin space and spin-degenerate bands, Eq.\
(\ref{eq:Jfk}) simplifies to the customary expression for Fermi's
golden rule. Therefore, Eq.\ (\ref{eq:FermiJfk}) can be viewed as
a generalization of Fermi's golden rule that explicitly takes into
account the spin degree of freedom. The commutator present in Eq.\
(\ref{eq:Fermi}) is a commutator in spin space that represents the
effect of spin precession due to spin-orbit interactions.

\subsection{Scattering term}

The scattering term $ \hat J (f_{{\bm k}})$ will now be evaluated
for an electron system with spin-orbit interactions. After
inserting a complete set of states, $\hat J (f_{{\bm k}})$ becomes
\footnote{In order to obtain Eq.\ (5) in the work of
Averkiev~\textit{et.~al.}, \cite{ave02} it is sufficient to
replace the EFA Hamiltonian $H_{{\bm k}}$ by the kinetic energy
term only, thus $H_{{\bm k}} = \hbar^2k^2/2m \equiv
\varepsilon_0$, and carry out the integration over time. We note
that a scattering term equivalent to that derived here was found
using the Keldysh formalism by E.~L.~Ivchenko \textit{et~al.},
Sov.~Phys.~JETP \textbf{71}(3), 551 (1990).}

\begin{widetext}
\begin{equation}
\hat J (f_{{\bm k}}) = \frac{n_i}{\hbar^2}\,
\lim_{\eta \rightarrow 0}\int_0^\infty dt' e^{-\eta t'} \sum_{{\bm
k}'} \left[ U_{{\bm k}{\bm k}'} e^{- i H_{{\bm k}'}
t'} (U_{{\bm k}'{\bm k}} f_{{\bm k}} - f_{{\bm k}'} U_{{\bm
k}'{\bm k}})\, e^{ i H_{{\bm k}} t'} - e^{- i H_{{\bm k}} t'}
(U_{{\bm k}{\bm k}'} f_{{\bm k}'} - f_{{\bm k}} U_{{\bm k}{\bm
k}'}) e^{ i H_{{\bm k}'} t'} U_{{\bm k}'{\bm k}} \right].
\end{equation}
The EFA Hamiltonian contains a kinetic energy term and a spin-orbit
coupling term, $H_{{\bm k}} = H_{{\bm k}}^\mathrm{kin} + H_{{\bm
k}}^\mathrm{so}$. In spin-1/2 electron systems, band structure
spin-orbit coupling can always be represented as a Zeeman-like
interaction of the spin with a wave vector-dependent effective
magnetic field ${\bm \Omega}_{\bm k}$, thus $H_{{\bm k}}^\mathrm{so}
= (1/2) \,{\bm \sigma}\cdot {\bm\Omega}_{\bm k}$. Common examples of
effective fields are the Rashba spin-orbit interaction, \cite{ras84}
which is often dominant in quantum wells with inversion asymmetry,
and the Dresselhaus spin-orbit interaction, \cite{dre55} which is
due to the inversion asymmetry of the underlying crystal lattice.
The spin-orbit interaction for electrons is usually much smaller
than the kinetic energy at typical Fermi energies, with the result
that terms that are second order in the ratio of the two, $H_{{\bm
k}}^\mathrm{so}/H_{{\bm k}}^\mathrm{kin}$, can usually be ignored.
In the basis in spin space spanned by spin eigenstates
$\ket{\uparrow}$ and $\ket{\downarrow}$ (commonly referred to as the
Pauli basis), taking into account the fact that $U_{{\bm k}{\bm
k}'}$ is diagonal in spin space, the scattering term simplifies to
\begin{equation}
\hat J (f_{{\bm k}}) =
\frac{n_iV_d}{\hbar^2}\,\lim_{\eta \rightarrow 0}\int_0^\infty dt'
e^{-\eta t'} \int\frac{d^dk'}{(2\pi)^d} W_{{\bm k}{\bm k}'} \,
\left[ e^{- i \hat H_{{\bm k}'} t'} \big( f_{{\bm k}} - f_{{\bm
k}'} \big) \, e^{ i H_{{\bm k}} t'}
 + e^{- i H_{{\bm k}} t'} \big( f_{{\bm k}} - f_{{\bm k}'} \big) e^{ i H_{{\bm k}'} t'}
 \right].
\end{equation}
Here $W_{{\bm k}{\bm k}'}= |U_{{\bm k}{\bm k}'}|^2$ is the
transition rate, and sums over wave vector have been converted
into integrals following the standard procedure $\sum_{{\bm k}'}
\rightarrow V_d \int d^dk'/(2\pi^d)$, where $d$ is the dimensionality of
the system and the normalization volume $V_d$ is chosen to be the
$d$-dimensional unit cell volume. The density matrix $f_{\bm k}$
is decomposed into a scalar part and a spin-dependent part,
$f_{\bm k} = n_{\bm k} \,\openone + S_{\bm k}$. Performing the
time integral, after a series of lengthy but straightforward
calculations, which are summarized in Appendix B, the scattering
term can be expressed in the form $(\hat{J}_0 + \hat{J}_s)\,
(n_{\bm k}) + \hat{J}_0(S_{\bm k})$, with
\begin{subequations}
\label{eq:scattering}
\begin{eqnarray}
\hat{J}_0\, (X_{\bm k}) & = & \frac{\pi n_iV_d}{2\hbar}\,
\int\frac{d^dk'}{(2\pi)^d}\, W_{{\bm k}{\bm k}'} (X_{\bm k} -
X_{{\bm k}'}) \left[ \delta(\epsilon_{{\bm k}+} - \epsilon'_{{\bm
k}+}) + \delta(\epsilon_{{\bm k}-} - \epsilon'_{{\bm k}-}) +
\delta(\epsilon_{{\bm k}+} - \epsilon'_{{\bm k}-}) +
\delta(\epsilon_{{\bm k}-} - \epsilon'_{{\bm k}+}) \right],
\hspace{2em} \\ [3ex] \label{eq:scatteringb} \hat{J}_s \, (X_{\bm
k}) & = & \frac{\pi n_iV_d}{2\hbar}\, \int\frac{d^dk'}{(2\pi)^d}\,
W_{{\bm k}{\bm k}'}(X_{\bm k} - X_{{\bm k}'}) \, {\bm
\sigma}\cdot(\hat{\bm \Omega}_{{\bm k}} + \hat{\bm \Omega}_{{\bm
k}'}) \left [\delta(\epsilon_{{\bm k}+} - \epsilon'_{{\bm k}+}) -
\delta(\epsilon_{{\bm k}-} - \epsilon'_{{\bm k}-}) \right]
\nonumber \\ [1ex] & & {} + \frac{\pi n_iV_d}{2\hbar}\,
\int\frac{d^dk'}{(2\pi)^d}\, W_{{\bm k}{\bm k}'}(X_{\bm k} -
X_{{\bm k}'}) \, {\bm \sigma}\cdot (\hat{\bm \Omega}_{{\bm k}} -
\hat{\bm \Omega}_{{\bm k}'}) \left[\delta(\epsilon_{{\bm k}+} -
\epsilon'_{{\bm k}-}) - \delta(\epsilon_{{\bm k}-} -
\epsilon'_{{\bm k}+}) \right].
\end{eqnarray}
\end{subequations}
\end{widetext}
$X_{\bm k}$ represents either $n_{\bm k}$ or $S_{\bm k}$, and
$\hat{\bm{\Omega}}_{{\bm k}}$ is a unit vector along
$\bm{\Omega}_{{\bm k}}$. The energies $\epsilon_{{\bm k}\pm}$
represent the eigenvalues of $H_{\bm k}$, and are given by
$\epsilon_{{\bm k}\pm} = \varepsilon_{0{\bm k}} \pm \Omega_{\bm
k}/2$, where the kinetic energy $\varepsilon_{0{\bm k}} = \hbar^2
k^2/2m^*$ and $\Omega_{\bm k} = |{\bm \Omega}_{\bm k}|$. The term
$\hat{J}_s \, (X_{\bm k})$ in Eq.\ (\ref{eq:scatteringb})
illustrates the fact that, when spin-orbit interactions are
present in the band structure, even a spin-independent scattering
potential usually gives rise to spin-dependent scattering in the
scattering integral.

\subsection{Time evolution of the density matrix in an external electric field}

In the presence of a constant uniform electric field ${\bm E}$,
${\bm k} = \bm{q} - e \bm{E} t/\hbar$ is the gauge-invariant
crystal momentum (with ${\bm q}$ the canonical momentum.) The
states $\ket{{\bm k} s}$ are chosen to have the form $\ket{{\bm k}
s} = e^{i \bm{q} \cdot \bm{r}} \ket{u_{\bm{k} s}}$, where
$\ket{u_{\bm{k} s}}$ are lattice-periodic functions. We subdivide
$f_{\bm{k}} = f_{0 \bm{k}} + f_{E {\bm{k}}} $, where the
equilibrium density matrix $f_{0 \bm{k}}$ is given by the
Fermi-Dirac distribution, and the correction $f_{E {\bm{k}}}$ is
due to the electric field ${\bm E}$. To first order in ${\bm E}$,
the correction $f_{E {\bm{k}}}$ satisfies
\begin{equation}\label{eq:boltze}
\pd{f_{E {\bm{k}}}}{t} + \frac{i}{\hbar}\, [H, f_{E {\bm{k}}}] +
\hat{J}\, (f_{E {\bm{k}}}) = \frac{e{\bm
E}}{\hbar}\cdot\pd{f_{0{\bm k}}}{{\bm k}}.
\end{equation}
The matrix $f_{E {\bm{k}}}$ is in turn divided, as above, into a
scalar part and a spin-dependent part, $f_{E {\bm{k}}} = n_{E
{\bm{k}}} \openone + S_{E {\bm{k}}}$. To first order in
$H^\mathrm{so}_{\bm k}/H^\mathrm{kin}_{\bm k}$, the scattering
term can be expressed as $ \hat J \, (f_{E {\bm{k}}}) = (\hat{J}_0
+ \hat{J}_s)\, (n_{E {\bm{k}}}) + \hat{J}_0\, (S_{E {\bm{k}}})$,
where the scattering operators have been defined in Eq.\
(\ref{eq:scattering}).

The driving term arising from the electric field in Eq.\
(\ref{eq:boltze}) is $(e{\bm E}/\hbar) \cdot \partial f_{0{\bm
k}}/\partial {\bm k}$. The equilibrium density matrix, $f_{0{\bm
k}}$, is subdivided as $f_{0{\bm k}} = n_{0{\bm k}}\openone +
S_{0{\bm k}}$, with a corresponding subdivision for the driving
term. The equation for $n_{E{\bm k}}$ is
\begin{equation}\label{eq:nE}
\pd{n_{E{\bm k}}}{t} + \hat J_0 \, (n_{E {\bm{k}}}) = \frac{e{\bm
E}}{\hbar}\cdot\pd{n_{0{\bm k}}}{{\bm k}}.
\end{equation}
The solution of this equation is given by the well-known
expression
\begin{equation}
n_{E{\bm k}} = \frac{e{\bm
E}\tau_p}{\hbar}\cdot\pd{n_{0{\bm k}}}{{\bm k}},
\end{equation}
in other words, $n_{E{\bm k}}$ describes the shift of the Fermi
sphere in the presence of the electric field ${\bm E}$. The
expression for the momentum relaxation time $\tau_p$ is a little
different depending on the dimensionality of the system. In three
dimensions, using $\gamma$ to denote the relative angle between
${\bm k}$ and ${\bm k}'$,
\begin{subequations}\label{eq:taup}
\begin{equation}
\frac{1}{\tau_p^{d=3}} = \frac{mkV_{d=3}}{2\pi \hbar^3} \int_0^\pi
d\gamma \, \sin\gamma \, W_{\bm{k}\bm{k}'} \, (1 - \cos\gamma).
\end{equation}
In two dimensions, with the same notation for $\gamma$,
\begin{equation}
\frac{1}{\tau_p^{d=2}} = \frac{mV_{d=2}}{2\pi \hbar^3}
\int_{0}^{2\pi}d\gamma \, W_{\bm{k}\bm{k}'} \, (1 - \cos\gamma).
\end{equation}
\end{subequations}

The spin-dependent part of the nonequilibrium correction to the
density matrix $S_{E{\bm k}}$ is interpreted as the spin density
induced by ${\bm E}$. The equation governing the time evolution of
$S_{E{\bm k}}$ is
\begin{equation}\label{eq:SBoltz}
\pd{S_{E {\bm{k}}}}{t} + \frac{i}{\hbar}\, [H_{\bm
k}, S_{E {\bm{k}}}] + \hat J_0 \, (S_{E {\bm{k}}}) =
\frac{e{\bm E}}{\hbar}\cdot\pd{S_{0{\bm k}}}{{\bm
k}} - \hat J_s \, (n_{E {\bm{k}}}).
\end{equation}
It is seen from Eq.\ (\ref{eq:SBoltz}) that spin-dependent
scattering gives rise to a renormalization of the driving term in
the equation for $S_{E{\bm k}}$. Evidently, this renormalization
has no analog in charge transport, see Eq.\ (\ref{eq:nE}).

So far, our work has not been restricted to the steady state. In
this context, we remark briefly that Eq.\ (\ref{eq:SBoltz}) can be
used to describe spin relaxation, and is valid both in the
presence and in the absence of external electric fields.
\cite{ave02} (In the presence of electric fields $S_{E {\bm{k}}}$
is an electric-field-induced nonequilibrium correction, whereas in
their absence it is to be interpreted more generally as a
nonequilibrium correction.)

\section{Steady states for conserved and precessing spins}

In the presence of band structure spin-orbit interactions, an
electron spin at wave vector ${\bm k}$ precesses about an
effective magnetic field ${\bm \Omega}_{\bm k}$. The spin can be
resolved into components parallel and perpendicular to ${\bm
\Omega}_{\bm k}$. In the course of spin precession the component
of the spin parallel to ${\bm \Omega}_{\bm k}$ is conserved, while
the perpendicular component is continually changing. It will prove
useful in our analysis to divide the spin distribution into a part
representing conserved spin and a part representing precessing
spin. This is accomplished below.

\subsection{Distributions of conserved and precessing spins}

Firstly, the effective source term, which enters the RHS of Eq.\
(\ref{eq:SBoltz}), is divided into two parts, $(e{\bm
E}/\hbar)\cdot\partial S_{0{\bm k}}/\partial{\bm k} - \hat J_s \,
(n_{E {\bm{k}}}) = \Sigma_{E{\bm k}\|} + \Sigma_{E{\bm k}\perp}$.
Here, $\Sigma_{E{\bm k}\|}$ commutes with the spin-orbit
Hamiltonian and is given by $\Sigma_{E{\bm k}\|} = \alpha_{\bm k}
H^\mathrm{so}_{\bm k}$, where
\begin{equation}
\alpha_{\bm k} = \frac{\displaystyle\trace \left\{\left[\frac{e{\bm
E}}{\hbar}\cdot\pd{S_{0{\bm k}}}{{\bm k}} - \hat J_s \, (n_{E
{\bm{k}}}) \right] \, H^\mathrm{so}_{\bm k} \right\}}{\displaystyle\trace
[(H^\mathrm{so}_{\bm k})^2]},
\end{equation}
while $\Sigma_{E{\bm k}\perp}$ is the remainder. In matrix
language $\Sigma_{E{\bm k}\perp}$ is \emph{orthogonal} to the
spin-orbit Hamiltonian and thus $\trace (\Sigma_{E{\bm k}\perp}\,
H^\mathrm{so}_{\bm k}) = 0$. Projections onto and orthogonal to
$H^\mathrm{so}_{\bm k}$ are most easily carried out by defining
projectors $P_\|$ and $P_\perp$. The actions of these projectors
on the basis matrices $\sigma_i$ are given by
\begin{subequations} \label{eq:projop}
\begin{eqnarray}
P_\| \sigma_i & = &
\frac{2\Omega_{{\bm k}i}\,H_{{\bm k}}^\mathrm{so}}{\Omega_{\bm k}^2},
\\ [1ex]
P_\perp \sigma_x & = &
\frac{(\Omega_{{\bm k}y}^2 + \Omega_{{\bm k} z}^2)\sigma_x -
\Omega_{{\bm k}x}\Omega_{{\bm k }y}\sigma_y - \Omega_{{\bm
k}x}\Omega_{{\bm k}z}\sigma_z}{\Omega_{\bm k}^2},
\nonumber \\
\end{eqnarray}
\end{subequations}
and the actions of $P_\perp$ on the remaining basis matrices are
obtained by cyclic permutations.

Secondly, $S_{E {\bm{k}}}$ is likewise divided into two terms:
$S_{E {\bm{k}}\|}$, commuting with the spin-orbit Hamiltonian and
$S_{E {\bm{k}}\perp}$, orthogonal to it. It is helpful to think of
$S_{E {\bm{k}}\|}$ as the distribution of conserved spins. $S_{E
{\bm{k}}\perp}$ can be thought of as the distribution of
precessing spins. Equation (\ref{eq:SBoltz}) is divided into
separate equations for $S_{E {\bm{k}}\|}$ and $S_{E
{\bm{k}}\perp}$:
\begin{subequations}
\begin{eqnarray}
& & \pd{S_{E {\bm{k}}\|}}{t}  +  P_\| \hat{J}_0 \,
(S_{E {\bm{k}}}) = \Sigma_{E{\bm k}\|},
\label{eq:Sigmaparallel}
\\ [1ex]
& & \pd{S_{E {\bm{k}}\perp}}{t}  + \frac{i}{\hbar}\, [H_{\bm k},
S_{E {\bm{k}}\perp}]
 + P_\perp \hat{J}_0 \, (S_{E {\bm{k}}}) = \Sigma_{E{\bm k}\perp} \label{eq:Sigmaperp}.
\hspace{3em}
\end{eqnarray}
\end{subequations}
The absence of the commutator $[H_{\bm k}, S_{E
{\bm{k}}\|}] = 0$ in Eq.\ (\ref{eq:Sigmaparallel})
indicates the absence of spin precession, while the commutator
$[H_{\bm k}, S_{E {\bm{k}}\perp}]$ in Eq.\ (\ref{eq:Sigmaperp})
represents spin precession. In order to solve Eqs.\
(\ref{eq:Sigmaparallel}) and (\ref{eq:Sigmaperp}) for arbitrary
scattering, it is necessary to expand $S_{E {\bm{k}}\|}$
and $S_{E {\bm{k}}\perp}$ in the transition rate $W_{{\bm k}{\bm
k}'}$, as
\begin{subequations}
  \begin{eqnarray}
    S_{E {\bm{k}}\|} & = & S_{E
{\bm{k}}\|}^{(-1)} + S_{E {\bm{k}}\|}^{(0)} + S_{E
{\bm{k}}\|}^{(1)} + \mathcal{O} (W_{{\bm k}{\bm k}'}^2), \\
\label{eq:Sperpexp} S_{E {\bm{k}}\perp} & = & S_{E
{\bm{k}}\perp}^{(0)} + S_{E {\bm{k}}\perp}^{(1)} + \mathcal{O}
(W_{{\bm k}{\bm k}'}^2).
  \end{eqnarray}
\end{subequations}
This expansion is indeed an expansion in the parameter
$\hbar/(\Omega_{\bm{k}} \, \tau_p)$, a fact that can be most
clearly seen by examining Eq.\ (\ref{eq:Sperpexp}) and noting that
the calculation of each term involves integration over time, which
brings in a factor of $1/\Omega_{\bm{k}}$, and the action of $\hat
J_0$. This expansion is therefore most suited to systems in the
weak scattering regime. The expansion of $S_{E {\bm{k}}\|}$ starts
at order $-1$, a fact which can be understood by inspecting Eq.\
(\ref{eq:Sigmaparallel}). In the steady state the time derivative
drops out, and the operator $\hat{J}_0$ is first order in $W_{{\bm
k}{\bm k}'}$, while the right-hand side is independent of $W_{{\bm
k}{\bm k}'}$. As a result, the expansion of the solution must
start at order $-1$. We examine next Eq.\ (\ref{eq:Sigmaperp}) for
$S_{E {\bm{k}}\perp}$. Since $H_{\bm k}$ is independent of
$W_{{\bm k}{\bm k}'}$, and the right hand side is also independent
of $W_{{\bm k}{\bm k}'}$, the expansion of $S_{E {\bm{k}}\perp}$
must start at order zero.

Having divided the nonequilibrium correction to the spin density
matrix into a part due to conserved spin and one due to precessing
spin, we wish to determine the contributions these parts make to
spin densities \cite{vor79, lev85, ede90, aro91, cha01, ino03,
eng07, kato04b, gan04, sil04} and spin currents \cite{eng06,
dya71, sin04, ada05, shi06, ras04, zha05, wang06, she05, han06,
kha06, shy06, dam04, schl05, jin06, li04, ino04, sch02, dim05,
mal05, eng05, nik06, ble06, kro06, zar06, mis04} in the steady
state. The steady state density of spin component $\sigma$ is
found by taking the trace $\trace (\hat s^\sigma \, S_{E{\bm
k}})$, where $\hat s^\sigma = (\hbar/2) \, \sigma^\sigma$. The
steady-state spin current is found by taking the trace $\trace
(\hat{\mathcal{J}}^\sigma_{i} S_{E{\bm k}})$, where $
\hat{\mathcal{J}}^\sigma_{i} = \hbar k_j s^\sigma/m^* + (1/4\hbar)
\, \partial \Omega^\sigma/\partial k_j \openone$. (The scalar term
has zero expectation value.) Henceforth, for clarity and
definiteness, integrals over wave vectors will be represented as
two-dimensional. In the integrals below $\theta'$ refers to the
polar angle of ${\bm k}'$. The extension to three dimensions is
straightforward.

\subsection{Steady state for conserved spins}
We have shown that the expansion of $S_{E {\bm{k}}\|}$
begins at order $-1$, and we wish to find the first term in this
expansion. In the steady state, Eq.\ (\ref{eq:Sigmaparallel}) for
the first term in this expansion can be written as
\begin{equation}
P_\| \hat{J}_0 \, (S_{E {\bm{k}}}^{(-1)}) =
\Sigma_{E{\bm k}\|}.
\end{equation}
This equation can be recast as
\begin{equation} \label{eq:Sparminus1}
\frac{S_{E {\bm{k}}\|}^{(-1)}}{\tau_0} -
P_\| \hat{J}'_0 (S_{E {\bm{k}}\|}^{(-1)}) =
\Sigma_{E{\bm k}\|},
\end{equation}
where we have introduced
\begin{subequations}
\begin{eqnarray}
\tau_0 & = & \frac{m^*}{2\pi \hbar^3} \int d\theta' W_{{\bm k}{\bm
k}'},
\\ [2ex] \hat{J}'_0 S_{E {\bm{k}}\|}^{(-1)} & = &
\frac{m^*}{2\pi \hbar^3} \int d\theta' W_{{\bm k}{\bm k}'} \, S_{E
{\bm{k}'}\|}^{(-1)}.
\end{eqnarray}
\end{subequations}
$\tau_0$ is the quantum lifetime of the charge carriers, i.e., the
time between two consecutive scattering events. It differs from
the momentum scattering time of Eqs.\ (\ref{eq:taup}) because, for
nonisotropic scattering mechanisms, the information about the
initial momentum is not lost after time $\tau_0$. Equation
(\ref{eq:Sparminus1}) can be solved iteratively for any scattering
\begin{equation} \label{eq:Sparminus1sol}
\begin{array}[b]{rl}
S_{E {\bm{k}}\|}^{(-1)} =
& \displaystyle
\Sigma_{E{\bm k}\|} \tau_0 + P_\| \hat{J}'_0
(\Sigma_{E{\bm k}\|}) \tau_0^2
\\ [2ex] & \displaystyle {}
+ P_\| \hat{J}'_0 [P_\| \hat{J}'_0 (\Sigma_{E{\bm
k}\|})] \tau_0^3 + \ldots
\end{array}
\end{equation}
The equations for higher orders in $W_{{\bm k}{\bm k}'}$ are
easily deduced. However, the term of order $-1$ is by far the
dominant one in the weak momentum scattering regime and is
expected to be dominant over a wide range of strengths of the
scattering potential.

We examine more closely the nature of the steady state established
for conserved spins. It is evident that this steady state involves
no spin precession, and that the correction $S_{E {\bm{k}}\|}$
depends explicitly on the nonequilibrium shift in the Fermi
surface and diverges in the ballistic regime, as $\tau_0
\rightarrow \infty$. In addition, it is important to note that
scattering terms contain only the even function $ W_{{\bm k}{\bm
k}'}$. As a result, the correction $S_{E {\bm{k}}\|}$ does not
give rise to a spin current. Inspection of Eq.\
(\ref{eq:Sparminus1sol}) shows that integrals of the form
\begin{equation}\label{eq:Jsintegral}
\int d\theta \, \hat{\mathcal{J}}^\sigma_{i} \, S_{E
{\bm{k}}\|}
\end{equation}
contain an odd number of powers of ${\bm k}$ and are therefore
zero. Consequently, in the absence of impurity spin-orbit
interactions, the distribution of conserved spins can give no spin
current. It can, however, give rise to a nonequilibrium spin
density \cite{vor79, ivc78, lev85, ede90, aro91, cha01, ino03,
eng07, kato04b, gan04, sil04}, since integrals of the form
\begin{equation}\label{eq:sintegral}
\int d\theta \, \hat{s}^\sigma \, S_{E {\bm{k}}\|}
\end{equation}
contain an even number of powers of ${\bm k}$ and may be nonzero.

\subsection{Steady state for precessing spins}
The equations for the contributions to $S_{E {\bm{k}}\perp}$ of
orders zero and one in $W_{{\bm k}{\bm k}'}$ are
\begin{subequations}
\begin{eqnarray}
\pd{S_{E {\bm{k}}\perp}^{(0)}}{t} + \frac{i}{\hbar}\, [H_{\bm k},
S_{E {\bm{k}}\perp}^{(0)}] &=& \Sigma_{E{\bm k}\perp} + P_\perp
\hat{J}'_0 \, (S_{E {\bm{k}}\|}), \label{eq:dSperp0dt}
\hspace{2em}
\\ [1ex] \pd{S_{E {\bm{k}}\perp}^{(1)}}{t} + \frac{i}{\hbar}\,
[H_{\bm k}, S_{E {\bm{k}}\perp}^{(1)}] &=& P_\perp \hat{J}'_0 \,
(S_{E {\bm{k}}\perp}^{(0)}).
\end{eqnarray}
\end{subequations}
To solve the equation for $S_{E {\bm{k}}\perp}^{(0)}$ it is
easiest to go into the interaction picture, obtain an expression
for $S_{E {\bm{k}}\perp}^{(0)}$, and then transform back to the
Schr\"odinger picture. This procedure yields for $S_{E
{\bm{k}}\perp}^{(0)}$
\begin{equation} \label{eq:Sperp0}
S_{E {\bm{k}}\perp}^{(0)} =
\frac{1}{2} \, \frac{\hat{\bm\Omega}_{\bm k}\times [{\bm
\Sigma}_{E {\bm{k}}\perp} + P_{\perp} \hat{J}'_0 \, ({\bm S}_{E
{\bm{k}}\|})] \cdot{\bm \sigma}}{\Omega_{\bm k}/\hbar},
\end{equation}
where we have written $\Sigma_{E {\bm{k}}\perp} = (1/2) \, {\bm
\Sigma}_{E {\bm{k}}\perp}\cdot{\bm \sigma}$ and $S_{E {\bm{k}}\|}
= (1/2) \, {\bm S}_{E {\bm{k}}\|}\cdot{\bm \sigma}$. The result
expressed by Eq.\ (\ref{eq:Sperp0}) is valid for any elastic
scattering. Since it does not depend explicitly on the form of the
impurity potential, this term is usually regarded as intrinsic.
Terms of higher order in $W_{{\bm k}{\bm k}'}$ are regarded as
extrinsic because they depend explicitly on the form of the
impurity potential. Nevertheless, it is evident from our work
that, if ${\bm \Sigma}_{E {\bm{k}}\perp} + P_{\perp} \hat{J}'_0 \,
({\bm S}_{E {\bm{k}}\|}) $ vanishes, then $S_{E
{\bm{k}}\perp}^{(0)}$ vanishes and all the terms in $S_{E
{\bm{k}}}$ of higher order in $W_{{\bm k}{\bm k}'}$ also vanish.

The steady state established for precessing spins is finite in the
clean limit. An argument similar to that given above for the
distribution of conserved spin $S_{E {\bm{k}}\|}$ shows that $S_{E
{\bm{k}}\perp}^{(0)}$ cannot lead to a nonequilibrium spin density
(although, as will be shown below, higher-order terms in $S_{E
{\bm{k}}\perp}$ can contribute to the spin density). For, taking
the expectation value of the spin operator, one arrives at
integrals of the form $\int d\theta \, \hat{s}^\sigma \, S_{E
{\bm{k}}\perp}^{(0)}$, which involve odd numbers of powers of
${\bm k}$ and are therefore zero. This term in the distribution of
precessing spin does, however, give rise to nonzero spin currents,
since integrals if the form $\int d\theta \,
\hat{\mathcal{J}}^\sigma_{i} \, S_{E {\bm{k}}\perp}$ contain an
even numbers of powers of ${\bm k}$ and may be nonzero.
Consequently, in the absence of spin-orbit coupling in the
scattering potential, nonequilibrium spin currents arise from spin
precession.

In order to investigate terms of higher order in $W_{{\bm k}{\bm
k}'}$, it is easiest to examine a concrete case. Although the
general conclusions apply to any elastic, spin-independent
scattering, this analysis will be done in the next section in the
context of short-range impurity scattering, where an exact
solution is possible, which reveals an interesting physical
picture.

\subsection{Short-range impurities}
An enlightening closed-form solution can be found for $S_{E
{\bm{k}}\|}$ and $S_{E {\bm{k}}\perp}$ for short-range
impurities. In this case $W_{{\bm k}{\bm k}'} \equiv W$ is a
constant and $\hat{J}_0\, S_{E {\bm{k}}} = (S_{E {\bm{k}}} -
\bar{S}_{E {\bm{k}}})/\tau_p$, where the scattering time
$\tau_p = \hbar^3/(mW)$ and the bar represents averaging over
directions in ${\bm k}$. Equation\ (\ref{eq:Sparminus1sol}) yields
a closed-form solution for $S_{E {\bm{k}}\|}$
\begin{equation}\label{eq:Sparallel}
S_{E {\bm{k}}\|} = \Sigma_{E {\bm{k}}\|} \, \tau_p +
P_\| \, (1 - \bar{P}_\|)^{-1} \bar{\Sigma}_{E
{\bm{k}}\|} \, \tau_p.
\end{equation}
This solution enters Eq.\ (\ref{eq:Sperp0}) for $S_{E
{\bm{k}}\perp}^{(0)}$. The equations for the contributions to
$S_{E {\bm{k}}\perp}$ of higher orders in $W$ can be easily
determined
\begin{subequations}
\begin{eqnarray}
S_{E {\bm{k}}\perp}^{(1)} & = & \frac{1}{2} \,
\frac{\hat{\bm\Omega}_{\bm k}\times \{ \hat{\bm\Omega}_{\bm k}
\times [{\bm \Sigma}_{E {\bm{k}}\perp} - \hat{J}_0 \, (S_{E
{\bm{k}}\|})] \} \cdot{\bm \sigma}}{\Omega_{\bm k}^2 \tau_p},
\\ [1ex]
S_{E {\bm{k}}\perp}^{(2)} & = & \frac{1}{2} \,
\frac{\hat{\bm\Omega}_{\bm k}\times\{\hat{\bm\Omega}_{\bm k}\times
\{\hat{\bm\Omega}_{\bm k}\times [{\bm \Sigma}_{E {\bm{k}}\perp} -
\hat{J}_0 \, (S_{E {\bm{k}}\|})]\}\} \cdot{\bm \sigma}}{\Omega_{\bm
k}^3 \tau_p^2} ,
\nonumber \\
\end{eqnarray}
\end{subequations}
and so on. Since $\Sigma_{E{\bm k}\perp}$ is orthogonal to the
spin-orbit Hamiltonian $H^\mathrm{so}_{\bm k}$, we have $\trace
(\Sigma_{E{\bm k}\perp} \, H^\mathrm{so}_{\bm k}) = 0$, which
tells us immediately that ${\bm \Sigma}_{E{\bm k}\perp} \perp
\hat{\bm \Omega}_{\bm k}$. As a result, $\hat{\bm\Omega}_{\bm
k}\times\hat{\bm\Omega}_{\bm k}\times{\bm \Sigma}_{E{\bm k}\perp}
= - {\bm \Sigma}_{E{\bm k}\perp}$, and
\begin{subequations}
\begin{eqnarray}
S_{E {\bm{k}}\perp}^{(1)} & = &  -
\frac{1}{2} \, \frac{[{\bm \Sigma}_{E {\bm{k}}\perp} - \hat{J}_0
\, (S_{E {\bm{k}}\|})] \cdot{\bm \sigma}}{\Omega_{\bm k}^2
\tau_p/\hbar^2},
\\ [1ex]
S_{E {\bm{k}}\perp}^{(2)} & = & -
\frac{1}{2} \, \frac{\hat{\bm\Omega}_{\bm k}\times [{\bm
\Sigma}_{E {\bm{k}}\perp} - \hat{J}_0 \, (S_{E
{\bm{k}}\|})] \cdot{\bm \sigma}}{\Omega_{\bm k}^3
\tau_p^2/\hbar^3}.
\end{eqnarray}
\end{subequations}
It is evident that the terms $S_{E {\bm{k}}\perp}^{(odd)}$ and
$S_{E {\bm{k}}\perp}^{(even)}$ give two separate geometric
progressions. These progressions are easily summed to give for
$S_{E {\bm{k}}\perp}$
\begin{equation} \label{eq:Sperp} \arraycolsep 0.3ex
\begin{array}[b]{rl}
\displaystyle S_{E {\bm{k}}\perp} = & \displaystyle
\frac{\bm{\Omega}_{\bm k} \times (\bm{\Sigma}_{E {\bm{k}}\perp}
\tau_p + P_\perp\, \bar{\bm S}_{E {\bm{k}}\|}) \cdot
\bm{\sigma}\, \tau_p}{2\hbar(1 + \Omega_{\bm k}^2 \tau_p^2/\hbar^2)} \\
[3ex] & \displaystyle {} - \frac{(\Sigma_{E{\bm k}\perp} \tau_p +
P_\perp\, \bar{S}_{E {\bm{k}}\|})}{1 + \Omega_{\bm
k}^2 \tau_p^2/\hbar^2}.
\end{array}
\end{equation}
Once again, if $\Sigma_{E{\bm k}\perp} \,\tau_p + P_\perp\,
\bar{S}_{E {\bm{k}}\|}$ vanishes, then all the corrections to
$S_{E {\bm{k}}\perp}$ of order zero and higher also vanish.

Let us analyze the two terms in $S_{E {\bm{k}}\perp}$. By
identifying terms in $S_{E {\bm{k}}\perp}$ even and odd in ${\bm
k}$, as was done above, it is evident that the first term on the RHS
of Eq.\ (\ref{eq:Sperp}) leads to a spin current, but no spin
density. The second term does not lead to a spin current, but it
does produce a spin density. Nevertheless, this term tends to zero
in the ballistic regime as well as in the strong momentum scattering
regime, and it is not expected to be dominant.

The closed-form solution found in this section shows that, in the
absence of spin-orbit interactions in the impurity potential,
there is only \emph{one} spin current, which in the weak
momentum-scattering limit is independent of $\tau_p$ and in the
strong momentum-scattering limit is $\propto \tau_p^2$. Bearing in
mind that if $S_{E {\bm{k}}\perp}^{(0)}$ vanishes all corrections
of higher order also vanish, we conclude that, in the absence of
spin-orbit interactions in the impurity potential, the distinction
between intrinsic (disorder-independent) and extrinsic
(disorder-dependent) spin currents is not useful.

\subsection{Steady state spin densities and currents}

Our work helps to understand the origins of nonequilibrium spin
densities and spin currents in electric fields. The preceding
sections illustrate the fact that nonequilibrium spin densities
have two origins. The first, giving the dominant contribution,
arises from the absence of precession. As charge carriers are
accelerated, a fraction of their spin is conserved and produces a
steady-state spin density, \cite{vor79, ivc78, lev85, aro91,
ede90, cha01, ino03, eng07, kato04b, gan04, sil04} a process which
has no analog in charge transport. Thus the dominant contribution
to the nonequilibrium spin density in an electric field exists
because in the course of spin precession a component of each
individual spin is preserved. For an electron with wave vector
$\bm k$, this spin component is parallel to ${\bm \Omega}_{\bm
k}$. In equilibrium the average of these conserved components is
zero. However, when an electric field is applied, the Fermi
surface is shifted, and the average of the conserved spin
components may be nonzero, as illustrated in Fig.~1. This
intuitive physical argument, to our knowledge absent to date from
the literature, explains why the nonequilibrium spin density
$\propto \tau_p^{-1}$ and \emph{requires} scattering to balance
the drift of the Fermi surface. It is interesting to note, also,
that, although spin densities in electric fields require the
presence of band structure spin-orbit interactions and therefore
spin precession, the dominant contribution arises as a result of
the absence of spin precession.

An additional contribution, which vanishes in both the ballistic
and the weak momentum scattering regimes, is associated with spin
precession. This contribution arises from the term on the last
line of Eq. (\ref{eq:Sperp}). The origin of this term can be
understood by noting that, in the presence of an electric field,
the effective magnetic field about which a spin precesses changes
slowly. \cite{sin04} (This is true in between scattering events.)
This change is contained in the gauge-invariant crystal wave
vector ${\bm k} = \bm{q} - e \bm{E} t/\hbar$. The fact that the
effective magnetic field is changing slowly causes the spin to
acquire a small component in the direction in which the effective
magnetic field is changing. This component is proportional to the
rate of change of the effective magnetic field and therefore, in
our case, to ${\bm E}$. It is associated with the `flow' of ${\bm
\Omega}_{\bm k}$ around the Fermi surface discussed by Shytov {\it
et al}. \cite{shy06} This argument explains why this term in the
spin density vanishes in the clean limit as well as in the strong
momentum scattering limit. For, in the absence of scattering, as
${\bm k}$ changes, the effective magnetic field will circle around
the Fermi surface, and the component of the spin following it will
average to zero. In the strong momentum scattering limit, on the
other hand, the spin will not have time to acquire a component in
the direction in which the effective magnetic field is changing,
due to the high frequency of scattering events.

\begin{figure}[tbp]
 \includegraphics[width=0.95\columnwidth]{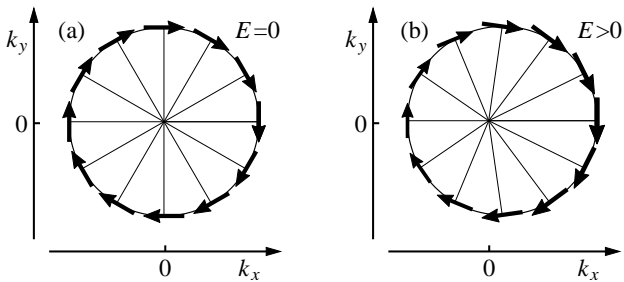}
 \hfill
 \caption{\label{fig:sia_e} Effective field ${\bm \Omega}_{\bm k}$
 at the Fermi energy in the Rashba model \cite{ras84} (a) without
 \cite{win04} ($E=0$) and (b) with an external electric field ($E>0$).}
\end{figure}

Furthermore, in the absence of spin-orbit interactions in the
impurity potentials, spin currents are associated with
displacement of spins. The relation between spin currents and spin
precession was made explicit in the work of Sinova \textit{et~al.}
\cite{sin04}

We remark that both $S_{E {\bm{k}}\|}$ and $S_{E
{\bm{k}}\perp}$ are invariant under time-reversal. As a result,
the tensor characterizing the response of spin currents to
electric fields is invariant under time reversal, whereas the
tensor characterizing the response of spin densities to electric
fields changes sign under time reversal, as expected in both
cases. \cite{eng06, dim05a}

\subsection{Interplay of conserved and precessing spin densities}

It is enlightening to compare the results obtained in the absence
of scattering (i.e., the clean limit) with the results obtained
when scattering is present. The aim is to obtain an understanding
of the way scattering processes affect steady-state spin
distributions in electric fields. This is done by comparing
results obtained using the approach outlined in this paper with
results obtained previously using Green's functions approaches,
both for the case when scattering is not included and for the case
in which scattering is taken into account. This process will aid
us in identifying the information contained in vertex corrections
to spin-related quantities in the framework of Green's
functions-based theories of systems with spin-orbit interactions.
The nature of this information is by no means obvious, and we will
show that it has no analog in charge transport.

In the absence of scattering, Eq.\ (\ref{eq:SBoltz}) takes the
form
\begin{equation}\label{eq:cleanSBoltz}
\arraycolsep 0.3ex
\begin{array}{rl}
\displaystyle \pd{S_{E {\bm{k}}}}{t} + \frac{i}{\hbar}\, [H_{\bm
k}, S_{E {\bm{k}}}] = &\displaystyle \frac{e{\bm
E}}{\hbar}\cdot\pd{S_{0{\bm k}}}{{\bm k}}.
\end{array}
\end{equation}
Comparison of Eqs.\ (\ref{eq:SBoltz}) and (\ref{eq:cleanSBoltz})
shows that the driving term in the equation for the nonequilibrium
spin distribution $S_{E {\bm{k}}}$ is renormalized by the term
$\hat J_s \, (n_{E {\bm{k}}})$, which accounts for spin-dependent
scattering.

In addition, Eq.\ (\ref{eq:dSperp0dt}) shows that scattering mixes
the distributions of conserved and precessing spins. This is so
because when one spin at wave vector ${\bm k}$ and precessing
about ${\bm \Omega}_{\bm k}$ is scattered to wave vector ${\bm
k}'$ and precesses about ${\bm \Omega}_{{\bm k}'}$, its conserved
component changes, a process which alters the distributions of
conserved and precessing spin. Consequently, scattering processes
in systems with spin-orbit interactions cause a renormalization of
the driving term for the spin distribution, contained in
$\Sigma_{E{\bm k}\perp}$, as well as scattering between the
conserved and precessing spin distributions, described by $P_\perp
\hat{J}'_0 \, (S_{E {\bm{k}}\|})$. Our analysis suggests that
contributions due to these two processes are contained in vertex
corrections to spin-dependent quantities found in Green's
functions formalisms.

In two dimensions, for Hamiltonians linear in wave vector, we find
that the renormalization term $\hat J_s \, (n_{E {\bm{k}}})$ does
not contribute to the spin current for any elastic scattering.
Therefore, the vertex correction to spin currents, found in other
work, \cite{sch02, mis04, ino04, ino03, dim05, kha06} represents
only scattering between conserved and precessing spin
distributions. By noting that in Eq.\ (\ref{eq:Sperp}) the
correction to the source term in the equation for the precessing
spin distribution has the form $P_\perp\, \bar{S}_{E {\bm{k}}\|}$
and thus depends on the steady state spin \emph{density}, it
becomes evident that the existence of a nonzero nonequilibrium
spin density does affect the spin current.

Furthermore, in three dimensions, for electrons in zincblende
crystals, which are described by the $k^3$-Dresselhaus model, we
find that, for short range impurities, the spin current obtained
after inclusion of scattering is the same as when scattering is
not included. It is known that the vertex correction to spin
currents also vanishes \cite{shy06} in these systems. Noting that
the steady-state spin density in zincblende crystals vanishes by
symmetry, and more generally vanishes in any non-gyrotropic
medium, \cite{ivc78} these observations reinforce our conclusion
relating to the connection between vertex corrections to spin
currents and the presence of a steady-state spin density. We
therefore expect vertex corrections to spin currents to vanish in
non-gyrotropic materials, in which no steady-state spin density is
possible in an electric field. \cite{ivc78}

\subsection{Comparison with previous work}

Our calculations for known cases give results in agreement with
previous work. Firstly, our results agree with previous calculations
of nonequilibrium spin densities. \cite{ede90, ino03} Furthermore,
in two dimensions, for Hamiltonians linear in wave vector, the spin
current vanishes for short-range impurities, \cite{sch02, mis04,
ino04, ino03, dim05, kha06} as well as for small-angle scattering.
\cite{kha06, shy06} (We find that in fact it vanishes for any
elastic scattering. \cite{sug06}) For spin-orbit Hamiltonians
characterized solely by one angular Fourier component $N$ (Ref.\
\onlinecite{shy06}) the spin current $\propto N$.

Spin currents in systems in which band structure spin-orbit
interactions are negligible were studied by Engel~
\textit{et~al.}, \cite{eng05} who demonstrated that spin currents
in those circumstances are due to skew scattering. Skew scattering
appears as a term of third order in the scattering potential
$U_{{\bm k}{\bm k}'}$ (which must include spin-orbit coupling
explicitly), whereas in this work we have restricted our
discussion to terms of second order in $U_{{\bm k}{\bm k}'}$, and
we have not considered higher-order skew-scattering effects.

\subsection{Observable effects}

Spin densities and spin currents excited by electric fields give
rise to observable effects. In materials in which a nonequilibrium
spin density is excited by an electric field, the presence of this
spin density can be observed, for example, by means of magnetic
circular dichroism \cite{sra06}, which has long been used as a
characterization tool for magnetic materials. Similarly, a spin
current flowing transversely to the direction of the charge
current (i.e., the spin-Hall effect) will give rise to a spin
accumulation at the edge of the sample. A spin current flowing
parallel to the direction of the charge current could be used as a
means of spin injection from one semiconductor into another, as
discussed in our recent work. \cite{dim07} Spin accumulation as a
result of a spin-Hall current, or a spin density injected by means
of a longitudinal spin current, can in turn be measured using
magnetic circular dichroism \cite{sra06} techniques developed
recently. We note that alternative techniques can be used to
observe nonequilibrium spin densities, \cite{vor79, kato04b, gan04,
sil04} spin currents, \cite{val06, liu06} and edge spin
accumulations. \cite{wun05, sih05, kato04}

\section{Summary}

We have demonstrated that, in the presence of band structure
spin-orbit interactions, the steady state established for the
carrier spin distribution contains two qualitatively distinct
contributions, corresponding to conserved and precessing spin. The
distribution of conserved spin acquires a nonequilibrium
correction that diverges in the ballistic regime. This correction
is responsible for the establishment of the dominant
nonequilibrium spin densities in electric fields. The distribution
of precessing spin acquires a nonequilibrium correction that is
finite in the ballistic regime. This correction is responsible for
the establishment of nonequilibrium spin currents in electric
fields and a small nonequilibrium spin polarization, which
vanishes in the ballistic and strong momentum scattering regimes.
We have demonstrated that, when spin-orbit interactions are
present in the band structure and absent from the impurity
potential, there is only one contribution to the spin current,
which appears independent of disorder in the ballistic regime but
dependent on disorder in the strong momentum scattering regime.
Moreover, we have also shown that scattering processes in systems
with spin-orbit interactions give rise to a renormalization of the
driving term in the equation for the spin distribution, as well as
scattering between the conserved and precessing spin
distributions, which sheds light on the nature of vertex
corrections in these systems.

The authors would like to acknowledge enlightening discussions
with E.~Rashba, Q.~Niu, A.~H.~MacDonald, J.~Sinova, D.~L.~Smith,
J.~Shi, H.~A.~Engel, and R.~A~Duine. The research at Argonne
National Laboratory was supported by the US Department of Energy,
Office of Science, Office of Basic Energy Sciences, under Contract
No. DE-AC02-06CH11357.

\appendix

\section{Matrix elements of a screened Coulomb potential} In two
dimensions, the matrix element $U_{{\bm k}{\bm k}'}$ of a screened
Coulomb potential between plane waves is given by
\begin{equation}
\arraycolsep 0.3ex
\begin{array}{rl}
\displaystyle U_{{\bm k}{\bm k}'} = & \displaystyle -
\frac{Ze^2}{\epsilon_0 V_{d=2}}\, \frac{1}{\sqrt{|{\bm k} - {\bm
k}'|^2 + 1/L_s^2}},
\end{array}
\end{equation}
where $Z$ is the ionic charge, $V_{d=2}$ corresponds to the unit
cell area, and $L_s$ is the screening length. The corresponding
expression in three dimensions is
\begin{equation}
\arraycolsep 0.3ex
\begin{array}{rl}
\displaystyle U_{{\bm k}{\bm k}'} = & \displaystyle -
\frac{Ze^2}{\epsilon_0 V_{d=3}}\, \frac{1}{|{\bm k} - {\bm k}'|^2
+ 1/L_s^2},
\end{array}
\end{equation}
where now $V_{d=3}$ is the unit cell volume.
\begin{widetext}
\section{Integrals over time}
For a Hamiltonian given by $H_{{\bm k}} = \varepsilon_{0{\bm k}}
\openone + (1/2) \, {\bm \sigma}\cdot {\bm\Omega}_{{\bm k}}$, the
product of two time evolution operators $e^{- i\hat H_{{\bm k}'}
t'} e^{ i \hat H_{{\bm k}}t'}$ can be written as
\begin{equation}
\arraycolsep 0.3ex
\begin{array}[b]{rl}
\displaystyle e^{- i\hat H_{{\bm k}'} t} e^{ i \hat H_{{\bm k}}t'}
= e^{ i(\varepsilon_0 - \varepsilon'_0) t'} \bigg[ & \displaystyle
\cos\frac{\Omega_{{\bm k}} t'}{2}\cos\frac{\Omega'_{{\bm k}}
t'}{2} - i\, {\bm \sigma}\cdot\hat{\bm \Omega}_{{\bm
k}'}\,\cos\frac{\Omega_{{\bm k}} t'}{2} \sin\frac{\Omega'_{{\bm
k}} t'}{2} + i\, {\bm \sigma}\cdot\hat{\bm \Omega}_{{\bm k}}\,
\sin\frac{\Omega_{{\bm k}} t'}{2}\cos\frac{\Omega'_{{\bm k}}
t'}{2} \\ [3ex] & \displaystyle {} + (\hat{\bm \Omega}_{{\bm
k}}\cdot\hat{\bm \Omega}_{{\bm k}'}+ i {\bm \sigma}\cdot\hat{\bm
\Omega}_{{\bm k}'}\times\hat{\bm \Omega}_{{\bm k}})\,
\sin\frac{\Omega_{{\bm k}} t'}{2}\, \sin\frac{\Omega'_{{\bm k}}
t'}{2} \bigg],
\end{array}
\end{equation}
with a similar expression holding for the Hermitian conjugate of
this product. In the time integrals the trigonometric functions
are expressed as complex exponentials, the integrals are
evaluated, and the results are replaced by their principal parts,
yielding
\begin{subequations}
\begin{eqnarray}
\int_0^\infty dt' \, e^{ i(\varepsilon_0 - \varepsilon'_0) t'}
\cos\frac{\Omega_{{\bm k}} t'}{2}\cos\frac{\Omega'_{{\bm k}}
t'}{2} & = & \frac{\pi\hbar}{4}\, [\delta(\epsilon_{{\bm k}+} -
\epsilon'_{{\bm k}+}) + \delta(\epsilon_{{\bm k}-} -
\epsilon'_{{\bm k}-}) + \delta(\epsilon_{{\bm k}+} -
\epsilon'_{{\bm k}-}) + \delta(\epsilon_{{\bm k}-} -
\epsilon'_{{\bm k}+})] \hspace{2em} \\ [1ex] \int_0^\infty dt' \,
e^{ i(\varepsilon_0 - \varepsilon'_0) t'} \cos\frac{\Omega_{{\bm
k}} t'}{2} \sin\frac{\Omega'_{{\bm k}} t'}{2}  & = &
\frac{\pi\hbar}{4i}\, [\delta(\epsilon_{{\bm k}+} -
\epsilon'_{{\bm k}-}) + \delta(\epsilon_{{\bm k}-} -
\epsilon'_{{\bm k}-}) - \delta(\epsilon_{{\bm k}+} -
\epsilon'_{{\bm k}+}) - \delta(\epsilon_{{\bm k}-} -
\epsilon'_{{\bm k}+})]
\\ [1ex]
\int_0^\infty dt' \,  e^{ i(\varepsilon_0 - \varepsilon'_0) t'}
\sin\frac{\Omega_{{\bm k}} t'}{2}\cos\frac{\Omega'_{{\bm k}}
t'}{2} & = & \frac{\pi\hbar}{4i}\, [\delta(\epsilon_{{\bm k}+} -
\epsilon'_{{\bm k}+}) + \delta(\epsilon_{{\bm k}+} -
\epsilon'_{{\bm k}-}) - \delta(\epsilon_{{\bm k}-} -
\epsilon'_{{\bm k}-}) - \delta(\epsilon_{{\bm k}-} -
\epsilon'_{{\bm k}+})]
\\ [1ex]
\int_0^\infty dt' \,  e^{ i(\varepsilon_0 -
\varepsilon'_0) t'} \sin\frac{\Omega_{{\bm k}} t'}{2}\,
\sin\frac{\Omega'_{{\bm k}} t'}{2} & = &
\frac{\pi\hbar}{4}\, [\delta(\epsilon_{{\bm k}+} - \epsilon'_{{\bm
k}-}) + \delta(\epsilon_{{\bm k}-} - \epsilon'_{{\bm k}+}) -
\delta(\epsilon_{{\bm k}-} - \epsilon'_{{\bm k}-}) -
\delta(\epsilon_{{\bm k}+} - \epsilon'_{{\bm k}+})].
\end{eqnarray}
\end{subequations}
Evaluation of all time integrals in this manner leads to Eq.\
(\ref{eq:scattering}) for the scattering term.
\end{widetext}

\end{document}